\documentclass[aps,prl,twocolumn,amsmath,amssymb,nofootinbib,pdffig,superscriptaddress]{revtex4-2}

\usepackage{graphicx}
\usepackage{dcolumn}
\usepackage{bm,psfrag}
\usepackage{dsfont}
\usepackage{float}
\usepackage{xcolor}
\usepackage{braket}

\usepackage{multirow}

\newcommand{\ben}{\begin{equation}}
\newcommand{\een}{\end{equation}}
\newcommand{\be}{\begin{equation}}
\newcommand{\ee}{\end{equation}}
\newcommand{\bea}{\begin{eqnarray}}
\newcommand{\eea}{\end{eqnarray}}
\newcommand{\ba}{\begin{eqnarray}}
\newcommand{\ea}{\end{eqnarray}}

\newcommand{\beq}{\begin{equation}}
\newcommand{\eeq}{\end{equation}}
\newcommand{\beqa}{\begin{eqnarray}}
\newcommand{\eeqa}{\end{eqnarray}}
\newcommand{\beqar}{\begin{eqnarray*}}
\newcommand{\eeqar}{\end{eqnarray*}}

\def\t6 {T_\mt{D6}}

\newcommand{\mt}[1]{\textrm{\tiny #1}}

\def\cale         {{\cal E}}

\def\ee           {{\rm e}}

\def\sqr#1#2{{\vcenter{\vbox{\hrule height.#2pt
 \hbox{\vrule width.#2pt height#1pt \kern#1pt
 \vrule width.#2pt}\hrule height.#2pt}}}}

\def\ee{\cale}

\def\aa1{\phi}
\def\cc1{\psi}

\def\ben{\begin{equation}}
\def\een{\end{equation}}
\def\bea{\begin{eqnarray}}
\def\eea{\end{eqnarray}}

\usepackage[bookmarks=false]{hyperref} 
\hypersetup{pdfstartview=FitH,pdfhighlight=/O,colorlinks=false}
\usepackage{cleveref}
\usepackage{orcidlink}

\begin{document}

\title{Boundary Condition dependent Universality Classes on a Hyperbolic Lattice}

\author{Pallabi Dey${}^{\orcidlink{0009-0007-2460-4545}}$}
\email{pallabi.dey@saha.ac.in}
\affiliation{Theory Division, Saha Institute of Nuclear Physics, 1/AF Bidhan Nagar, 
Kolkata 700064, India}
\affiliation{Homi Bhabha National Institute, Training School Complex, Anushaktinagar, 
Mumbai 400094,India}

\author{Debasish Banerjee${}^{\orcidlink{0000-0003-0244-4337}}$}
\email{D.Banerjee@soton.ac.uk}
\affiliation{School of Physics and Astronomy, University of Southampton, University Road, SO17 1BJ, UK}
\author{Arnab Kundu${}^{\orcidlink{0000-0002-1994-3346}}$}
\email{arnab.kundu@saha.ac.in}
\affiliation{Theory Division, Saha Institute of Nuclear Physics, 1/AF Bidhan Nagar, 
Kolkata 700064, India}

\author{Ritam Sinha${}^{\orcidlink{0000-0002-4217-9577}}$}
\email{ritamsinha.physics@gmail.com}
\affiliation{QuNu Labs Pvt.~Ltd.,~M.~G.~Road, Bengaluru 560025 Karnataka, India.}

\begin{abstract}
 We study the ferromagnetic Ising model on finite hyperbolic tessellations of Euclidean AdS$_2$ with open and 
 wired boundary conditions. Because a finite fraction of spins remains at the boundary as the lattice 
 grows, these conditions select distinct thermodynamic behaviours. In the $\{5,4\}$ tessellation, 
 Monte-Carlo simulations using efficient worm algorithms on open boundaries (OBC) yield a transition near 
 $\beta_c J=0.632(3)$, with susceptibility data collapsing under scaling by the total number of spins $N$. 
 The fitted finite-size exponents, $1/\bar{\nu} \simeq 0.162$ and 
 $\gamma/\bar{\nu} \simeq 0.743$, differ substantially from the mean-field volume scaling. Wired boundaries 
 (WBC), which correlate boundary spins, instead show a transition around $\beta_c J=0.348(4)$ to an 
 ordered phase consistent with the mean-field exponents. While the mean-field criticality with WBC is 
 consistent with previous studies and with the suppression of independent boundary fluctuations, 
 the OBC exponents hint at the presence of a new universality class. Results from other tessellations support 
 the robustness of the observed scaling with OBC. We discuss a possible route to interpolate between the
 different boundary conditions. Our findings show that boundary dynamics must be specified when 
 characterizing critical behaviour on hyperbolic lattices.
\end{abstract}

\maketitle

{\it Introduction:} Universality allows systems with different microscopic interactions 
to exhibit the same critical behaviour \cite{Cardy1996}. Finding a transition outside 
an established universality class, therefore, reveals which assumptions about the long-distance 
theory cease to apply. Geometry offers a way to test those assumptions: in particular, 
negative curvature introduces an intrinsic length scale that may influence critical behaviour. 
The approach to the thermodynamic limit in flat space is predominantly governed by the bulk degrees 
of freedom since the boundary is subleading. Negatively-curved hyperbolic geometries, on the 
other hand, are {\it non-amenable}, so that the number of sites in successive radial layers 
grows exponentially and, therefore, the boundary accounts for an ${\cal O}(1)$ fraction of the entire 
volume of the geometry. Consequently, it is expected that the allowed boundary dynamics 
will have a crucial impact on the thermodynamic limit of a spin system defined on a 
hyperbolic lattice \cite{Lyons2000, Jonasson1999RandomCluster}. 

\begin{figure}
    \centering
    \includegraphics[width=0.8\linewidth]{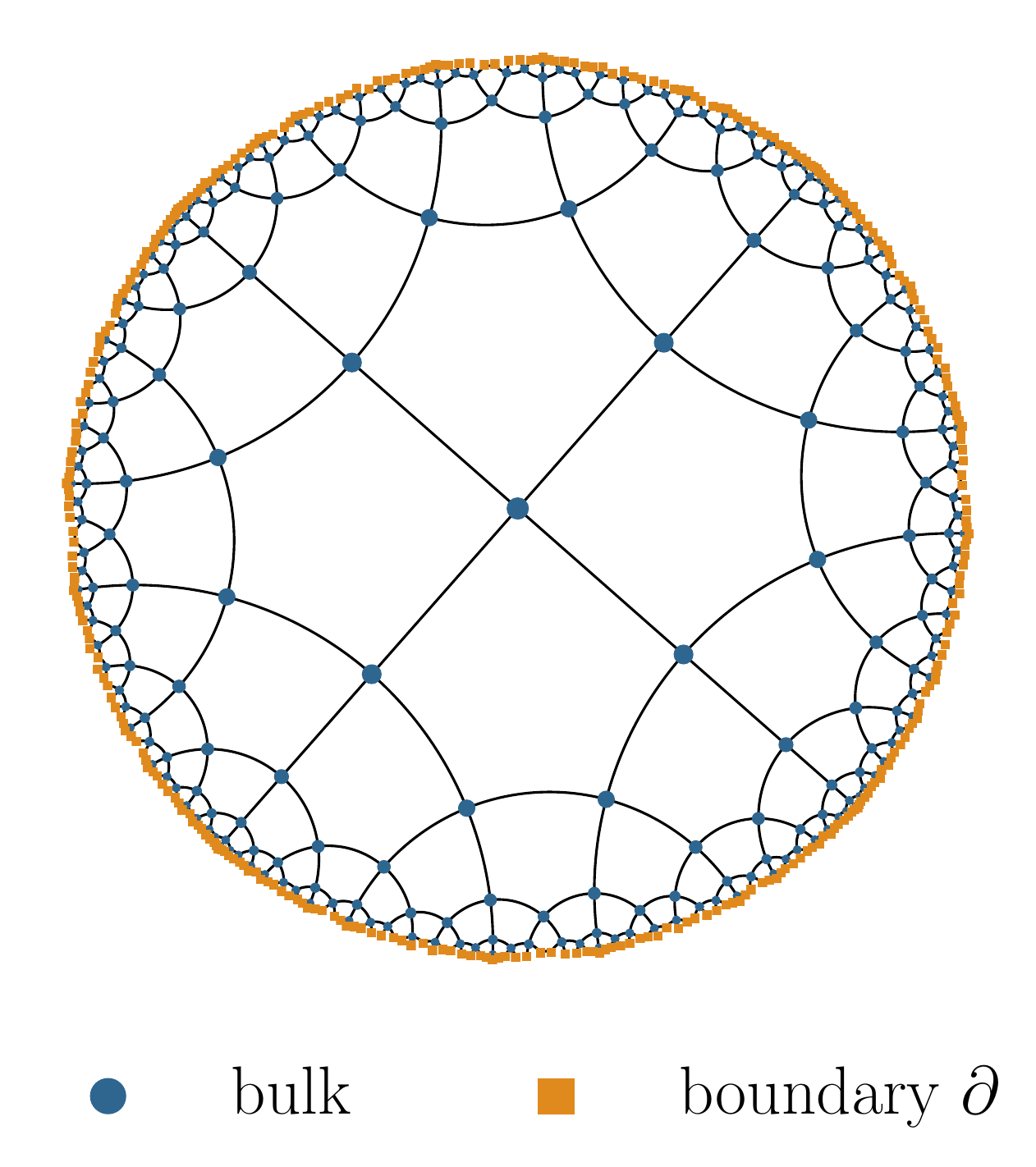}
    \caption{The $\{5,4\}$ hyperbolic tessellations with $N = 597$ sites in the Poincar\'{e} disk. }
    \label{fig:tiling54}
\end{figure}

 A continuum treatment would make the role of curvature particularly transparent, but 
nonperturbative calculations require a tractable regularization. Regular $\{p,q\}$ tessellations 
provide such a setting: they retain the geometry of the hyperbolic plane while permitting 
explicit spin-model simulations \cite{Brower2021,Asaduzzaman2022}. Increasing the number 
of radial layers at fixed $\{p,q\}$ probes the thermodynamic limit of a given lattice model
at fixed curvature in lattice units. 
Fig.~\ref{fig:tiling54} shows the finite $\{5,4\}$ tiling used for our principal numerical results.
Changing the tessellation then varies that curvature, 
allowing us to test whether the critical exponents persist across different negatively curved 
lattices. Because $\{p,q\}$ also changes local connectivity, agreement between tessellations 
provides evidence of robustness of the critical phenomena to curvature and local structure. 

The distinction between bulk and boundary behaviour has long complicated the interpretation 
of Ising models on negatively curved graphs. Eggarter \cite{Eggarter1974} showed on Cayley 
trees that a response deep in the interior can signal ordering even though the zero-field 
free energy of the full open tree remains analytic. Non mean-field estimates were obtained
using finite-size scaling (FSS) on hyperbolic tessellations \cite{Shima2006} with free 
boundaries, which drifted towards mean-field values when outer layers were excluded from 
the measured observables. 
Corner-transfer-matrix studies likewise found mean-field-like behaviour in the interior, with 
a finite geodesic correlation length at the transition \cite{Krcmar2008,Iharagi2010}. 
Investigations on periodic hyperbolic lattices (without an extensive open boundary), found 
mean-field critical exponents across different tessellations \cite{Breuckmann2020}. Thus, 
results for the full open system and for its interior or periodic counterpart need not 
characterize the same thermodynamic limit.

 The boundary has also been studied as a physical sector in its own right. Analyses of a 
hyperbolic graph with an exposed boundary found mean-field-like surface critical behaviour 
\cite{Baek2011}, while simulations on tessellations found power-law boundary correlations 
on both sides of the bulk transition \cite{Asaduzzaman2022}. More recently, two transitions 
under open boundary conditions (OBC), separated by a boundary-sensitive intermediate phase 
when the system is subject to infinitesimal (boundary) magnetic field were observed in 
\cite{f4sj-rwvj}; wiring the boundary (WBC) instead produced an ordering 
transition at the higher of the two temperatures. 

  We compare the critical scaling selected by these two boundary conditions, OBC and WBC,
for the same ferromagnetic Ising model on a $\{5,4\}$ tessellation. OBC produces a set of 
susceptibility scaling exponents markedly different from mean-field values. Similar anomalous 
exponents recur across tessellations with different curvatures, discussed in the companion 
paper \cite{Dey2026b} supporting the possibility of a common boundary-dependent universality 
class. WBCs instead yield an ordering consistent with mean-field exponents at a different 
critical coupling. The physical origin of this difference arises due to the different 
nature of the fluctuations involved in the two cases. With OBC, the extensive boundary retains 
its local fluctuations; with WBC, a common auxiliary spin introduces a collective boundary 
coupling. The resulting ensembles can have different thermodynamic limits.

\begin{figure}[!htpb]
    \centering
    \includegraphics[width=\linewidth]{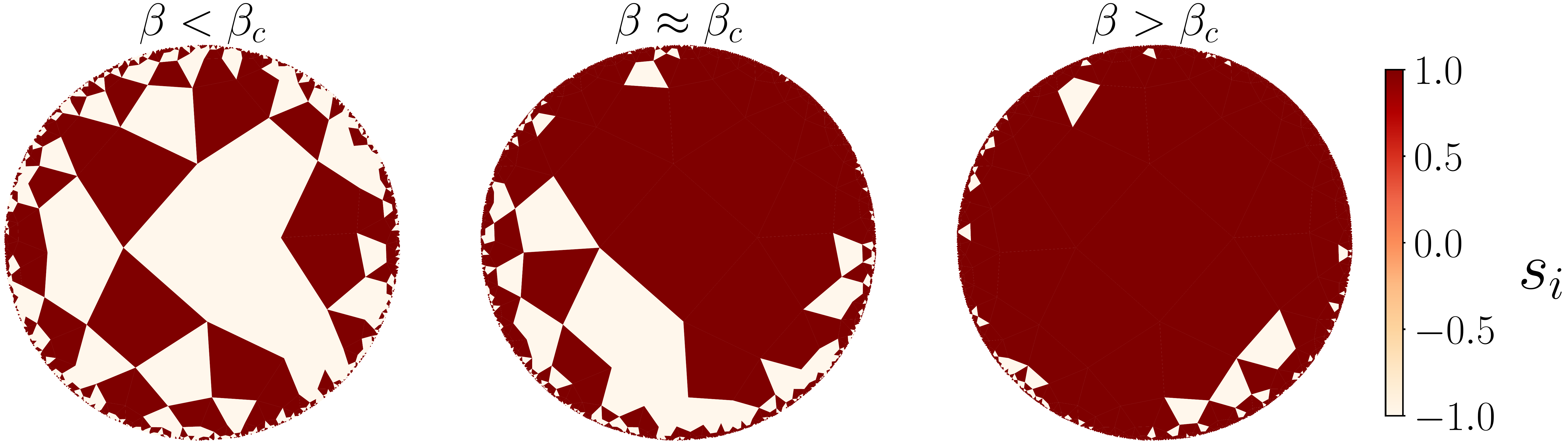}    
    \caption{Illustrative spin configuration of the Ising model on the $\{5,4\}$ 
    tessellations in the disordered phase $\beta < \beta_c$, near criticality 
    $\beta \approx \beta_c$ and ordered phase $\beta >\beta_c$. }
    \label{fig:snapshot_tile54}
\end{figure}

 The appropriate finite-size variable is itself part of this distinction. At fixed negative 
curvature, the number of spins increases exponentially with radial extent, $N\sim e^{cL}$, so 
the flat-lattice relation $N\sim L^d$ does not apply. Volume-based scaling has been 
considered before \cite{Kulkarni2022} for the ferromagnetic Ising model on random regular 
graphs to show that their finite-size scaling is governed by the variable $(T/T_c-1) N^{1/2}$ 
rather than a power of the graph diameter ($\sim \log N$). We also demonstrate scaling with $N$, 
but with anomalous powers, matching across four different tessellations, and providing a hint of 
a new universality with OBCs. Our results suggest that boundary conditions must be specified 
while discussing universality on non-amenable lattices. 

{\it The Model:} We consider finite, concentric patches of regular $\{p,q\}$ tessellations, 
in which $q$ regular $p$-gons meet at each interior vertex. The choice of $\{p,q\}$ fixes the 
curvature in lattice units, and we can define the following scaled curvature 
$K_{\rm scaled} = -\pi \left(1 - \frac{2}{p} - \frac{2}{q}\right)$. Our studies have 
considered four different tessellations $\{3,7\}$, $\{3,8\}$, $\{5,4\}$, and $\{6,6\}$
with curvatures $-\pi/21, -\pi/12, -\pi/10$ and $-\pi/3$ respectively.  
We show the appropriate figures and analysis for the $\{5,4\}$ tessellation, tabulate the critical 
exponents for all the tessellations studied, and present the detailed results 
in a companion paper \cite{Dey2026b}. All $\{p,q\}$ tessellations are generated with the \mbox{HYPERTILING} package \cite{SciPostPhysCodeb.34, SciPostPhysCodeb.34-r1.3}, which provides nearest-neighbors information required for our simulation.

 Each point on the vertex occupies an Ising spin, $s_i=\pm1$, with ferromagnetic 
nearest-neighbour interactions
\begin{equation}
  H_{\mathrm{OBC}}=-J\sum_{\braket{i,j}} s_i s_j,\qquad J>0,
\end{equation}
where $i$ and $j$ are nearest-neighbour sites with a fixed co-ordination number, $q$, except at 
the boundaries, where they are connected only to the spins in the same and in the inner layers.
The ferromagnetic coupling is set to $J=1$ throughout our work.
For wired boundary conditions, every boundary spin is coupled to a common auxiliary spin 
$s_{\mathrm{aux}} = \pm 1$:
\begin{equation}
 H_{\mathrm{WBC}}=H_{\mathrm{OBC}} - J_b s_{\mathrm{aux}}\sum_{i\in\partial V_N}s_i.
\end{equation}
The auxiliary spin is dynamical, so a simultaneous reversal of all physical spins and 
$s_{\mathrm{aux}}$ preserves the global $\mathbb Z_2$ symmetry. Moreover, we will consider
$J_b = J$ until the end, when we interpolate between the boundary conditions.
In both ensembles, $N$ counts physical lattice spins, with their magnetization defined as
\begin{equation}
  m=\frac{1}{N}\sum_{i=1}^{N}s_i.
\end{equation}
We measure the susceptibility and Binder cumulant as
\begin{equation}
\chi=N \braket{m^2}, \qquad U_2=1-\frac{\braket{m^4}}{3\braket{m^2}^2}.
\end{equation}

 We study both boundary conditions non-perturbatively using Markov-chain Monte Carlo methods. 
Single-spin Metropolis updates provide a direct simulation in the spin representation, while
the worm algorithm simulates the high-temperature expansion and substantially reduces the
autocorrelation times near criticality \cite{Prokof_ev_2001,Deng2007,Wolff2009}. 
Agreement between the two formulations provides an independent check of the simulations. 
Further algorithmic and simulation details are given in \cite{Dey2026b}.

\begin{figure}[H]
    \centering
    \includegraphics[width=0.95\linewidth]{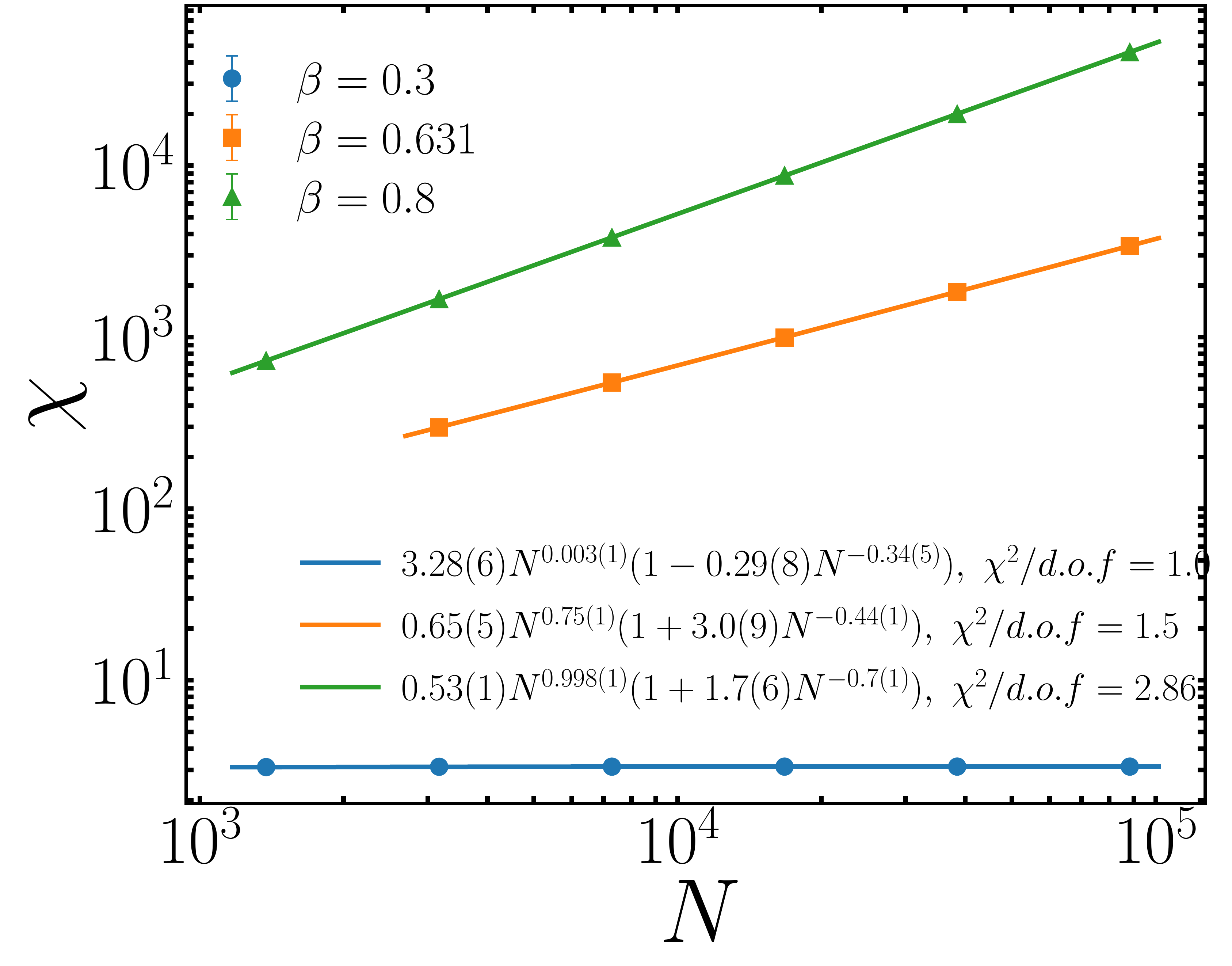}
    \caption{Volume dependence of the susceptibility $\chi$ on the $\{5, 4\}$ 
    tessellations with OBC in the disordered phase $\beta = 0.3$ ($< \beta_c$), 
    near criticality $\beta = 0.631$ ($\approx \beta_c$) and ordered phase 
    $\beta = 0.8$ ($>\beta_c$).}
    \label{fig:chi_with_N}
\end{figure}

{\it Critical behaviour:} We first establish that the OBC system exhibits the familiar 
thermal transition from a disordered to an ordered phase. Fig.~\ref{fig:chi_with_N} 
shows the volume dependence of the total susceptibility, $\chi$, at representative couplings 
below, near, and above the transition. In the disordered phase, $\beta<\beta_c$, 
$\chi$ saturates in the thermodynamic limit. 
In the ordered phase, $\beta>\beta_c$, the finite system samples the two 
$\mathbb Z_2$-related magnetized sectors, so that $\braket{m}=0$ while 
$\braket{m^2}$ approaches a nonzero value; consequently, $\chi=N \braket{m^2}$ 
grows extensively. Near the transition, however, the susceptibility follows a nontrivial intermediate power, 
$\chi \sim N^{y_\chi}$, with $y_\chi \simeq 0.75$ for the $\{5,4\}$ tessellation. This 
behaviour is clearly distinct from both saturation in the disordered phase and extensive 
growth in the ordered phase. It motivates identifying $y_\chi=\gamma/\bar{\nu}$ and 
undertaking a systematic finite-size-scaling analysis to determine $\beta_c$, $\bar{\nu}$, 
and $\gamma/\bar{\nu}$.

 {\it FSS with open boundaries:} On a flat Euclidean lattice, FSS is conventionally 
formulated in terms of its linear extent \cite{Cardy1996}. A finite hyperbolic tessellation 
instead has two natural measures of size: the total number of lattice sites, $N$, and the 
number of radial layers, $L \sim \log N$ (an exact relation between $N$ and $L$ is provided
in Tab.~\ref{tab:N_L}). Geometrically, $N$ measures the volume, and $L$ 
the radial geodesic extent. The curvature $K$, fixed by the choice of $\{p,q\}$, provides 
an additional intrinsic scale of the model. It is therefore not {\it a priori} evident 
whether conventional one-parameter scaling survives, or whether $L$ or $N$ is the appropriate 
finite-size variable for an observable. Empirically, our global observables (such as $\chi$) 
are accurately described by a one-parameter scaling form based on $N$, similar to that 
found for the Ising model on random regular graphs \cite{Kulkarni2022}. 
This precedent motivates the choice of $N$, but does not determine the scaling exponents, 
which must instead be extracted from the FSS fit. 
\begin{figure}[!tbh] 
    \centering
    \includegraphics[width=\linewidth]{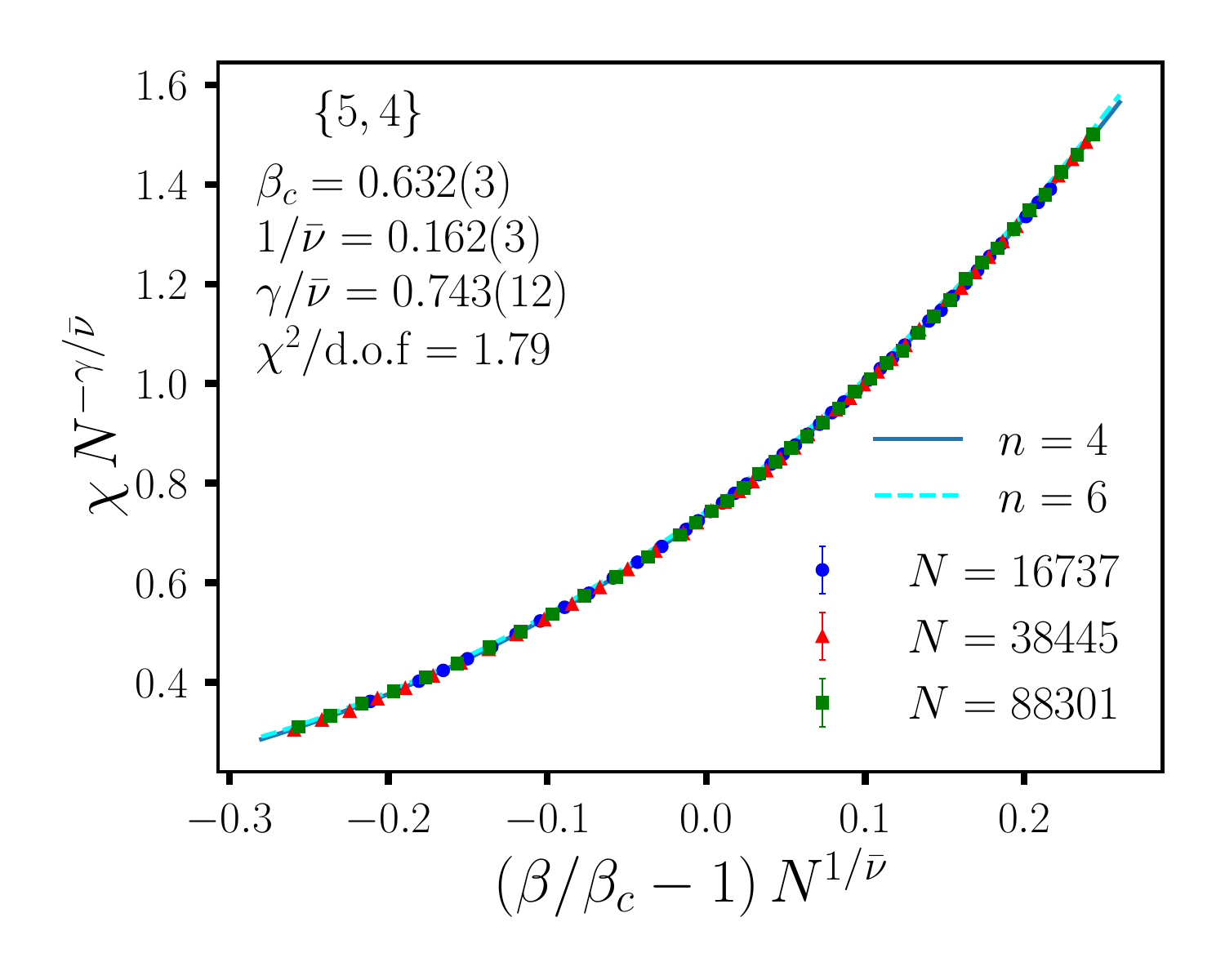}
    \includegraphics[width=0.92\linewidth]{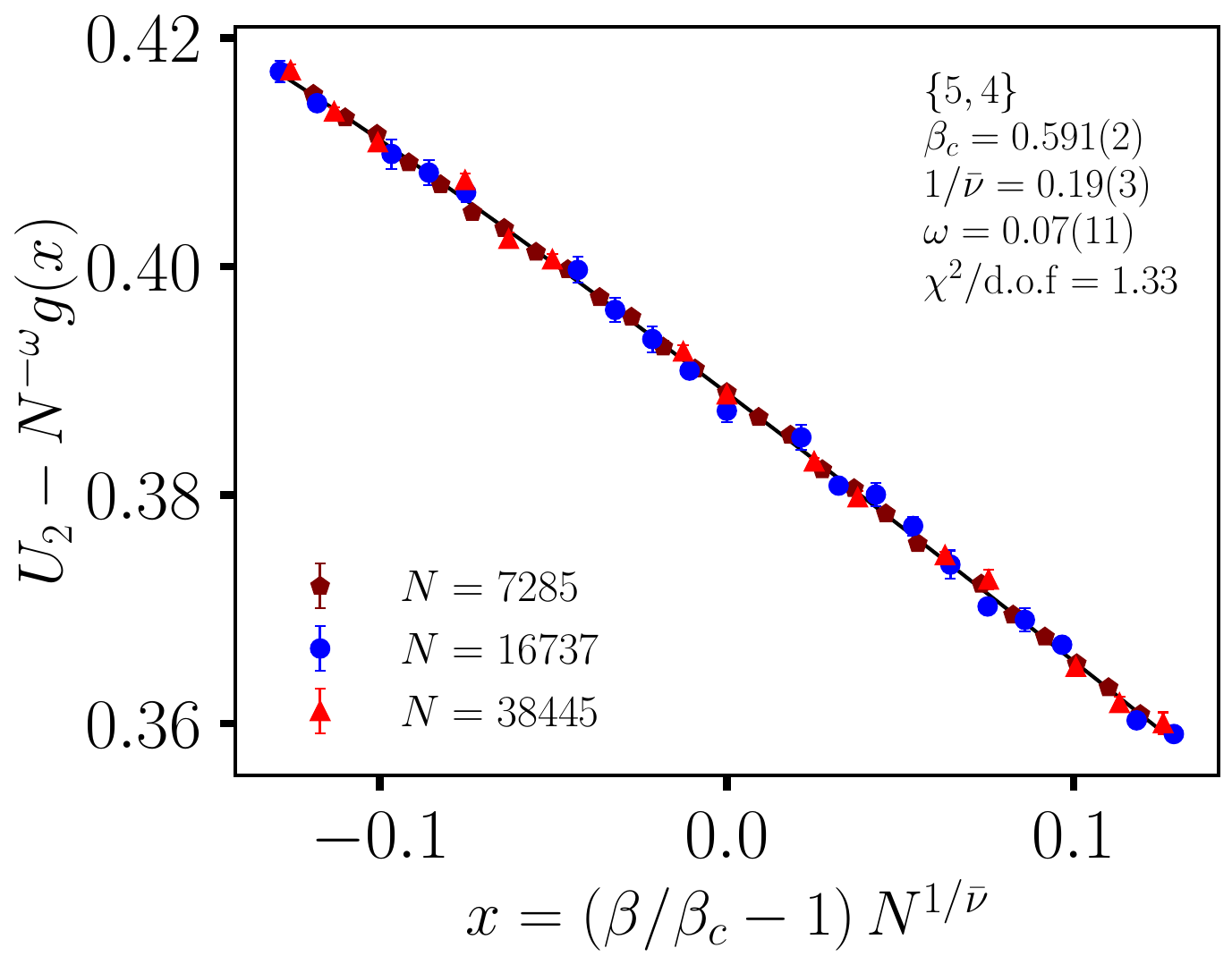}
     \caption{(Top) FSS collapse on the $\{5, 4\}$ tiling using the volume-scaling with
     OBC. Solid and dashed lines are fits to fourth and sixth order polynomials respectively. 
     Quoted exponents use the quartic polynomial fit. (Bottom) The FSS collapse of Binder 
     cumulant at the fixed $\beta_c = 0.591(2)$, obtained from Binder-crossing (see SM 
     Sec.~\hyperref[sec:FSS_OBC]{II}), but requires a scaling correction (in the same variable), 
     where $g(x)$ is a linear polynomial.}
     \label{fig:FSS54}
\end{figure}
 Universality would be reflected in scaling exponents that are insensitive to 
microscopic details of the lattice regularization. In the present setting, comparing 
different $\{p,q\}$ tessellations constitutes a particularly stringent test, since both
the local connectivity and the curvature in the lattice units change. It is therefore not 
{\it a priori} necessary that different tessellations share the same critical exponents. Nevertheless, we find that the exponents extracted with OBC are approximately independent 
of $\{p,q\}$ over the four tessellations studied. Within the finite-size uncertainties 
discussed below, this persistence supports their interpretation as a common set of universal 
exponents rather than tessellation-specific effective behaviour.

 To formulate the scaling hypothesis, we define $\gamma$ through the thermodynamic susceptibility,
and introduce the correlation-volume exponent $\bar{\nu}$ as follows,
\begin{equation}
 \chi_\infty(t)\sim |t|^{-\gamma}, \qquad N_\xi(t)\sim |t|^{-\bar{\nu}},
\end{equation}
where $N_\xi$ denotes the characteristic correlation volume and $\xi$ the correlation length. 
For a $d$-dimensional Euclidean 
system, $N_\xi \sim \xi^d$ and hence $\bar{\nu}=d\nu$. No such power-law relation between 
volume and geodesic correlation length exists on a hyperbolic lattice, and we define the 
corresponding FSS ansatz as
\begin{equation} \label{eq:FSS}
 \chi(t,N)=N^{\gamma/\bar{\nu}} F\left(tN^{1/\bar{\nu}}\right), \qquad
 t=\frac{\beta}{\beta_c}-1 .
\end{equation}
At criticality, this gives $\chi(0,N)\sim N^{\gamma/\bar{\nu}}$, directly connecting the 
scaling ansatz to the anomalous power observed in Fig.~\ref{fig:chi_with_N}. Fig.~\ref{fig:FSS54}
shows the FSS analysis of $\chi$ and the Binder $U_2$. We obtain $1/{\bar\nu}=0.162(3)$,
$\gamma/\bar{\nu}=0.743(12)$, and $\beta_c=0.632(3)$. 
The computed exponents are stable with respect to the fit range
and the fit functions. Moreover, it is also possible to fit the Binder cumulant $U_2$ 
and obtain the exponent $1/\bar{\nu} = 0.19(3)$ consistent with the estimate from $\chi$. 
However, the fit requires a scaling correction (with the exponent $\omega = 0.07(11)$) to
produce an acceptable $\chi^2$ and yields a $\beta_c = 0.591(2)$, approximately 7\% lower 
than the estimate from $\chi$ (see SM, Sec.~\hyperref[sec:FSS_OBC]{II}), indicating
substantial finite-size systematics associated with the smaller lattices accessible to
Binder analysis.

In addition, we have performed a detailed analysis on three other tessellations, which
will be reported in the companion paper \cite{Dey2026b}. The results on the critical 
exponents are summarized in Tab.~\ref{tab:fss_exponents}. 
While $\beta_c$ varies between tessellations, as expected, the exponents $\gamma/\bar{\nu}$ 
agree within 1-2$\sigma$ for all tessellations. The stability of $\gamma/\bar{\nu}$, 
together with the comparatively modest but statistically significant spread in 
$1/\bar{\nu}$, supports a possible common asymptotic scaling, although residual 
finite-size effects remain. The spread of the critical exponents is essentially due to
the larger finite-size effects. For the tessellation $\{5,4\}$, we clarify some
aspects in SM Sec.~\hyperref[sec:FSS_OBC]{II} for completeness, but more details of other tessellations are discussed in the companion paper. 
\begin{table*}
  \centering
  \setlength{\tabcolsep}{14pt}
  \renewcommand{\arraystretch}{1.4}
  \begin{tabular}{cccccc}
    \hline
    Lattice & Order($n$) & $\beta_c$ & $\gamma/\bar{\nu}$ & $1/\bar{\nu}$ & $\chi^2/\mathrm{d.o.f.}$ \\
    \hline
    \multirow{2}{*}{$\{3,7\}$} & 4th &  0.313(1) &  0.745(10)& 0.177(2) & 1.97 \\
                               & 6th &  0.313(1) &  0.747(10)& 0.177(2) & 1.97   \\
    \hline
    \multirow{2}{*}{$\{3,8\}$} & 4th &  0.315(1) &  0.73(1) &0.170(2)  &1.82  \\
                               & 6th &  0.315(1) &  0.73(1) &0.170(2)  & 1.82 \\
    \hline
    \multirow{2}{*}{$\{5,4\}$} & 4th &  0.632(3) &  0.743(12) &0.162(3)  &1.79  \\
                               & 6th &  0.631(3) &  0.741(12) &0.162(3)  & 1.81 \\
    \hline
    \multirow{2}{*}{$\{6,6\}$} & 4th & 0.723(2) & 0.738(3) &0.172(1)  & 2.34 \\
                               & 6th & 0.722(2) & 0.737(3) & 0.171(1) & 2.06 \\
    \hline
  \end{tabular}
  \caption{Finite-size scaling parameters for the $\{3,7\}$, $\{3,8\}$, $\{5,4\}$, and $\{6,6\}$ tilings,
           obtained from data collapse with 4th- and 6th-order scaling polynomials with OBC.}
  \label{tab:fss_exponents}
\end{table*}

\begin{figure}
   \centering
   \includegraphics[width=\linewidth]{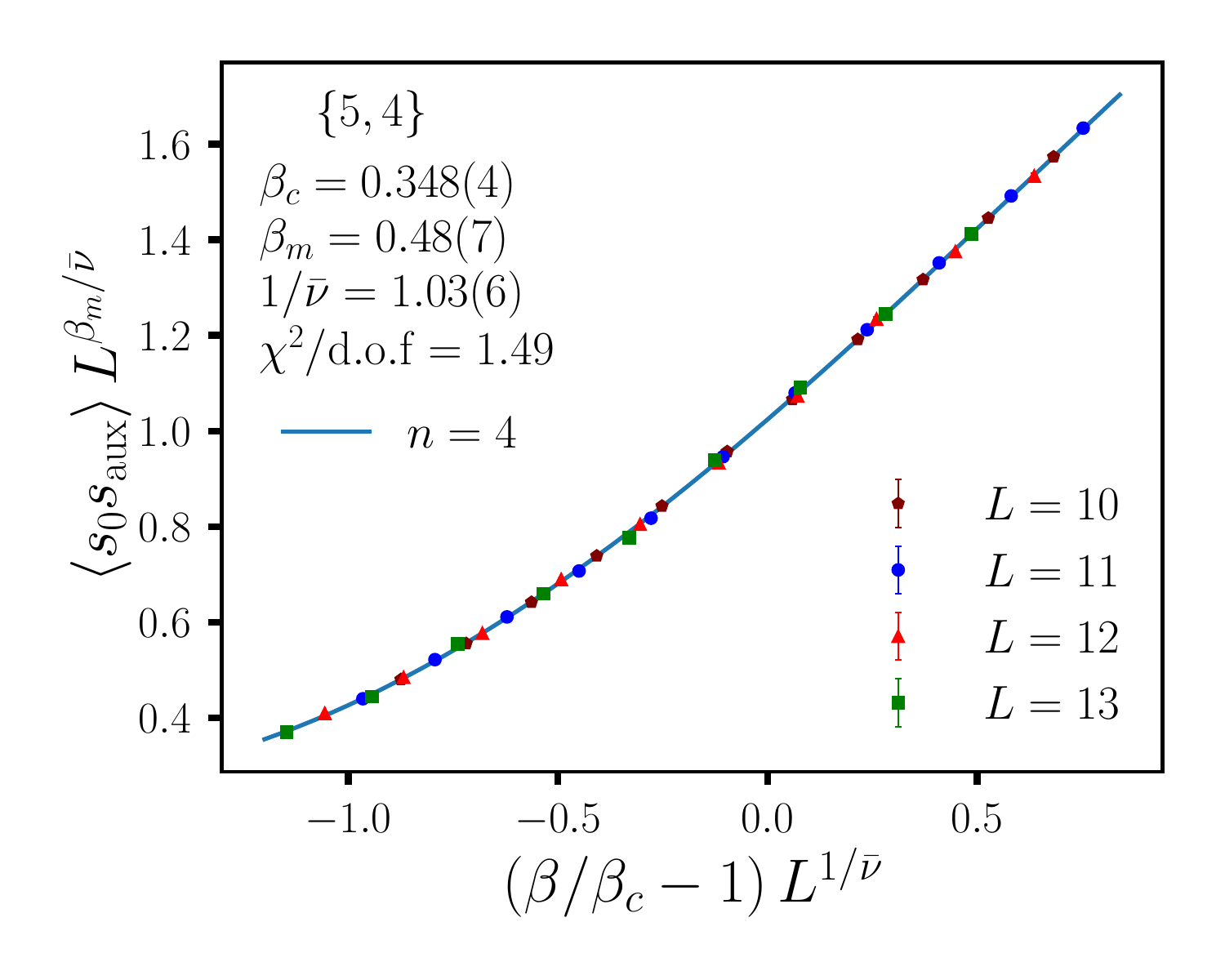}
   \caption{FSS fit to the two point correlation function of the spins separated by the largest
   geodesic distance on the $\{5,4\}$ show a scaling collapse with critical exponents that are
   consistent with mean-field scaling.}
   \label{fig:tile54_WBC}
\end{figure}

{\it FSS with wired boundaries:} With WBC, the auxiliary spin suppresses 
independent boundary fluctuations while retaining a coherent mode in which the boundary spins 
fluctuate collectively with $s_{\rm aux}$. The worm algorithm updates the auxiliary
spin frequently, in contrast to a local update. Consequently, 
the global magnetization is strongly influenced by the extensive boundary and is not the most 
direct probe of bulk ordering. We instead consider the symmetry-invariant central-to-boundary 
correlator at the largest possible geodesic distance,
\begin{equation}
 m_0^{\rm W}(L,\beta)=\braket{s_0 s_{\rm aux}},
\end{equation}
which measures whether the boundary order penetrates to the centre of the lattice. Since this
correlation spans the entire radial extent, we use the scaling form
\begin{equation}
 m_0^{\rm W}(L,\beta) = L^{-\beta_m/\bar{\nu}}\, \Phi\!\left[(\beta-\beta_c)L^{1/\bar{\nu}}\right].
\end{equation}
The fit, shown in Fig.~\ref{fig:tile54_WBC} yields $\beta_c=0.348(4)$, $\beta_m=0.48(7)$, and
$1/\bar{\nu}=1.03(6)$, consistent with the mean-field values $\beta_m=1/2$ and $1/\bar{\nu}=1$.
We also note that the value of $\beta_c$ is close to the estimate of \cite{f4sj-rwvj}
for this tessellation ($0.3573$). Additional evidence of the reliability of our
extraction of $\beta_c$ and the critical exponents is provided in SM Sec.~\hyperref[sec:WBC]{III}. 
Our analysis involving the Eggarter phase, as noted in \cite{f4sj-rwvj}, will be reported
in \cite{Dey2026b}.

\begin{figure}
    \centering
    \includegraphics[width=\linewidth]{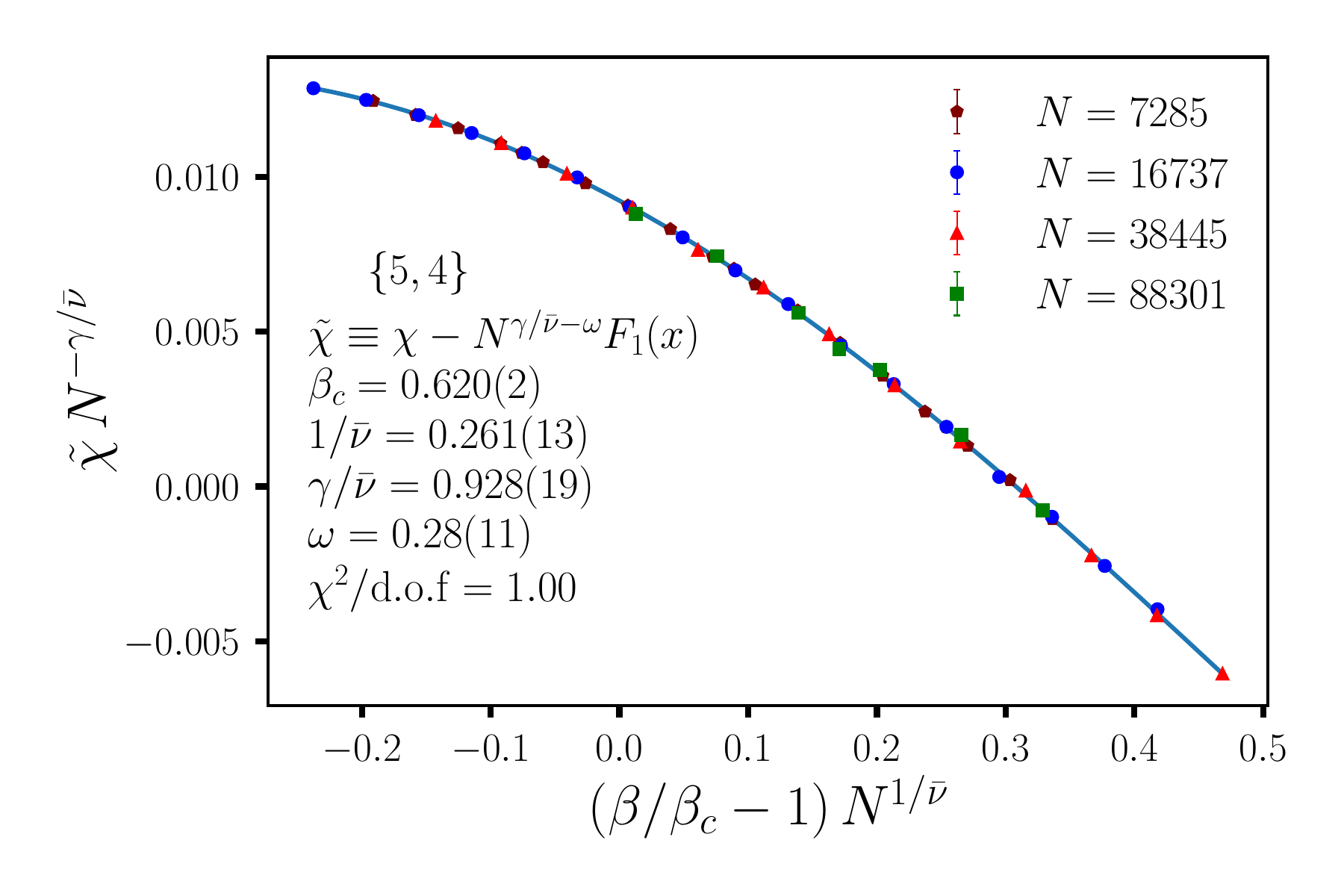}
    \caption{Finite-size scaling collapse of the (effective) scaling corrected 
    susceptibility $\tilde{\chi}$ on the $\{5, 4\}$ lattice under WBC, where 
    boundary spin couples to auxiliary spin with a layer-dependent coupling 
    $J_b  = 1/\partial V_N$. The solid line indicates the fourth order polynomial 
    fit and the correction function $F_1(x)$ is the quadratic polynomial in 
    $x = (\beta/\beta_c - 1) N^{1/\bar{\nu}}$.}
    \label{fig:FSS_L_dep}
\end{figure}

{\it FSS with layer-dependent coupling:} Finally, we have attempted an analysis
which can coherently control the boundary conditions and interpolate between OBC and WBC
studied earlier. To do this, we scale the coupling of each boundary spin to the auxiliary 
spin as $J_b=J/N_b$, where $N_b=|\partial V_N|$. Unlike a fixed $J_b$, for which the total 
boundary coupling grows with $N_b$, this choice keeps the coupling to a coherently aligned 
boundary configuration of order unity. As shown in Fig.~\ref{fig:FSS_L_dep}, a volume-based 
collapse including the leading correction to scaling gives $\beta_c=0.620(2)$, 
$1/\bar{\nu}=0.261(13)$, and $\gamma/\bar{\nu}=0.928(19)$, which lie between the effective 
OBC and fixed-WBC results. These values should presently be regarded as crossover exponents: 
the drift of $\beta_c$ towards its OBC value and the sizeable scaling correction favour an 
eventual flow to OBC, although the available sizes cannot exclude a distinct intermediate 
fixed point.

{\it Discussion and conclusions:}
We have shown that the critical behaviour of the Ising model on a hyperbolic lattice is
sensitive not only to the local Hamiltonian, but also to the thermodynamic boundary condition. 
With OBC, the $\chi$ exhibits volume-based FSS with non-mean-field exponents that are 
approximately stable across the four tessellations studied. With WBC, the auxiliary spin 
suppresses independent boundary fluctuations while retaining a coherent boundary mode. 
The central-to-boundary correlator $\braket{s_0 s_{\rm aux}}$ then obeys radial finite-size 
scaling consistent with mean-field criticality. The use of $N$ for the global OBC 
susceptibility and $L$ for the WBC two-point function reflects the different spatial 
character of these observables.

Hyperbolic lattices belong to the broader class of non-amenable graphs, for which 
boundary-dependent Gibbs states and phase transitions are known to occur \cite{Lyons2000}. 
Our results further suggest that the thermodynamic boundary condition can select 
the critical scaling itself. 
These findings support an extension of the conventional notion of universality on
non-amenable lattices: in addition to symmetry, dimensionality, and local degrees of
freedom, the allowed boundary dynamics may form part of the definition of the critical
theory. A corresponding renormalization-group description of the boundary flow remains
an important open problem.

\noindent {\it Acknowledgements:} A.K. acknowledges the support of the Humboldt Research 
Fellowship for Experienced Researchers by the Alexander von Humboldt Foundation and for 
the hospitality of Theoretical Physics III, Department of Physics and Astronomy,
Julius-Maximilians-Universit\"{a}t W\"{u}rzburg and the support from the ICTP through 
the Associates Programme (2024-2030) during the course of this work. RS acknowledges the 
support of the Royal Society-Newton International Fellowship NIF/R1/221054-Royal Society 
during the initial stages of the project. The affiliation for RS is listed solely for 
identification. The views expressed are entirely those of the authors and do not reflect 
the official position or strategic roadmap of QuNu Labs Pvt. Ltd. This work was conducted 
independently of RS’s employment, utilising no corporate funding, resources, or proprietary 
technology from QuNu Labs Pvt. Ltd. and the latter holds no intellectual property rights 
over this manuscript. D.B. would like to thank STFC (UK) consolidated grant ST/X000583/1 
and continued support from the Alexander von Humboldt Foundation (Germany) in the context 
of the research fellowship for experienced researchers. We thank computing resources of 
SINP to derive these results.
This work used the DiRAC@Durham facility managed by the Institute for Computational Cosmology 
on behalf of the STFC DiRAC HPC Facility as well. The equipment was funded by BEIS capital 
funding via STFC capital grants ST/P002293/1, ST/R002371/1 and ST/S002502/1, Durham University 
and STFC operations grant ST/R000832/1. This work was also performed using resources provided 
by the Cambridge Service for Data Driven Discovery (CSD3) operated by the University of 
Cambridge Research Computing Service, provided by Dell EMC and Intel using Tier-2 funding 
from the Engineering and Physical Sciences Research Council (capital grant EP/T022159/1), 
and DiRAC funding from the Science and Technology Facilities Council. 

\bibliography{ref} 
\appendix
\section{Supplementary Material} 
\label{suppmat} 
\subsection{I. Some notes on Hyperbolic Geometry}
\label{sec:Geometry}
Any hyperbolic geometry with $\{p,q\}$-tiling is characterised by its Gaussian curvature 
$K$ \cite{PhysRevLett.125.053901},
\begin{align}
    K_{\rm Gauss} = -\frac{\pi p }{A_{\rm poly}}\left(1-\frac{2}{p}-\frac{2}{q}\right)~~,
\end{align}
where $A_{\rm poly}$ is the area of each $p$-sided unit polygon in the lattice. Given 
that the thermodynamic limit of the hyperbolic lattice is independent of $A_{\rm poly}$, 
one can scale away the area by assuming $A_{\rm poly}=p$. The scaled Gaussian curvature 
$K_{\rm scaled}$ is therefore given only in terms of the tiling parameters $\{p,q\}$ \cite{f4sj-rwvj}:
\begin{align}
    K_{\rm scaled} = -\pi\left(1-\frac{2}{p}-\frac{2}{q}\right)~~.
\end{align}
Note that the curvature is symmetric in $p$ and $q$.
For flat-space tilings $\{4,4\}$, $\{3,6\}$, $\{6,3\}$ etc. $K=0$, where the hyperbolic 
tilings reduce to a Euclidean manifold.
Another important asymptotic limit is the $p\rightarrow\infty$, where the hyperbolic lattice 
reduces to the Cayley tree. In this limit, the Gaussian curvature is strictly negative for 
any $q>2$, and forms a wide class of negatively curved lattices with no closed loops. The 
Ising model on the Cayley tree shows a transition from the paramagnetic to the Eggarter phase, 
at the phase transition point $K_{\rm c, Eggarter}$. However, it is assumed that the transition 
from the Eggarter to the ferromagnetic phase cannot be observed in this setting due to the 
absence of closed loops.

For a finite hyperbolic tessellation $\{p,q\}$, a direct consequence of $K_{\rm scaled} < 0$ 
is that the number of sites $N$ grows exponentially with the number of layers $L$.
The total number of sites $N$ for each layer $L$ of the $\{5, 4\}$ tessellation is listed 
in \cref{tab:N_L}.
\begin{table}[H]
    \centering
    \setlength{\tabcolsep}{14pt}
  \renewcommand{\arraystretch}{1.4}
    \begin{tabular}{ccc}
    $\{p,q\}$ & layer & total \# of sites \\
        &  L & N \\
        \hline
        \hline       
       &  8 & 1377  \\
      & 9 & 3169 \\
      & 10 & 7285 \\
      \{5,4\}   & 11 & 16737 \\
      & 12 & 38445 \\
      & 13 &  88301 \\
      \hline
    \end{tabular}
    \caption{Exponential growth of total number of sites $N$ with each layer L of the $\{5,4\}$ tessellation.}
    \label{tab:N_L}
\end{table}

\subsection{II. Finite size correction with OBC}
\label{sec:FSS_OBC}
\begin{figure}
    \centering
    \includegraphics[width=\linewidth]{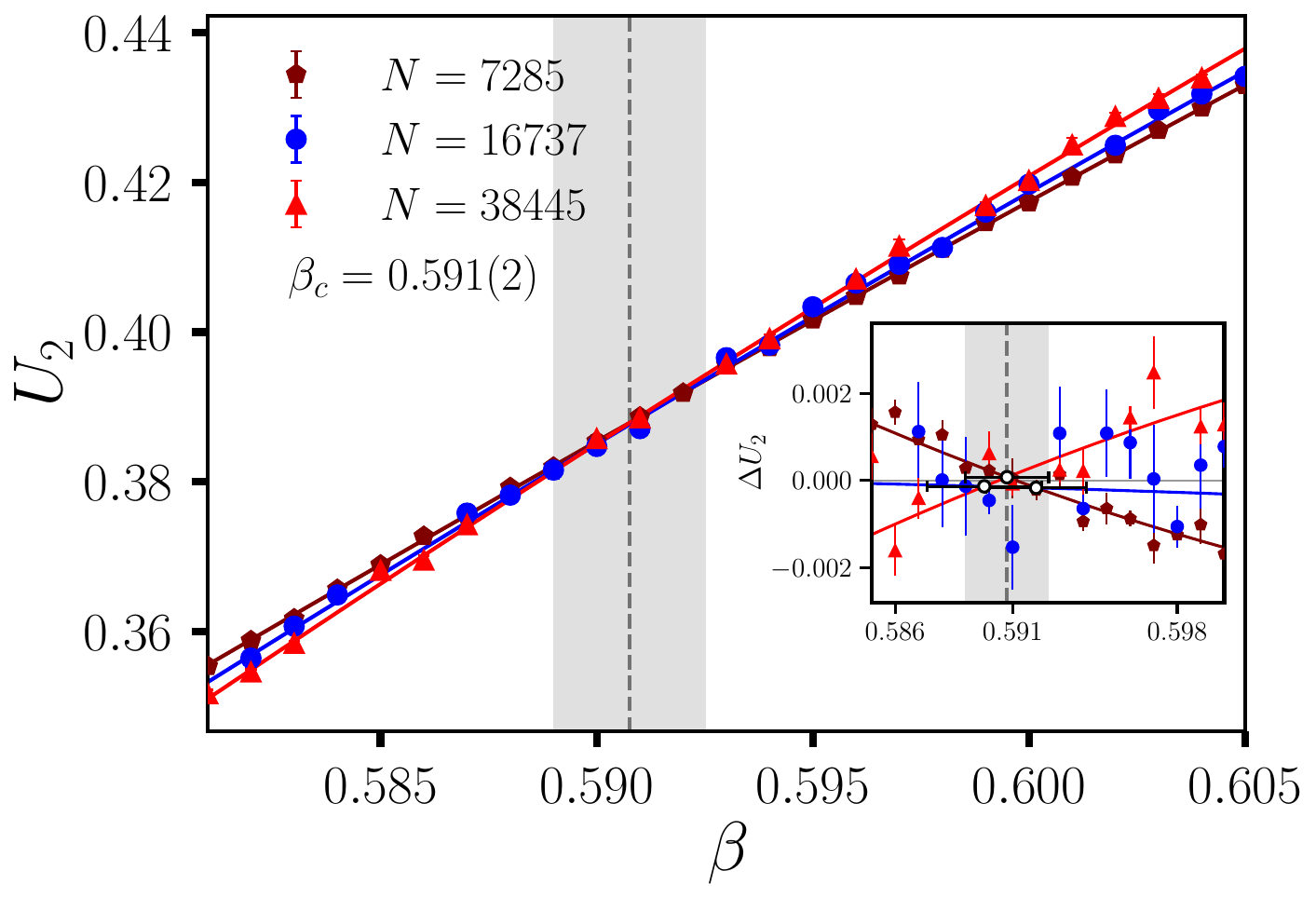}    
    \caption{Estimate of the (pseudo-)critical point from the crossing of Binder 
    cumulant curves on the $\{5,4\}$ tessellation. Inset: $\Delta U_2 = U_2 - \braket{U_2}$, 
    the deviation from the average of the fitted curves, which makes the intersection clearly 
    visible.}
    \label{fig:crossing_binder}
\end{figure}

\begin{figure}
    \centering
    \includegraphics[width=0.9\linewidth]{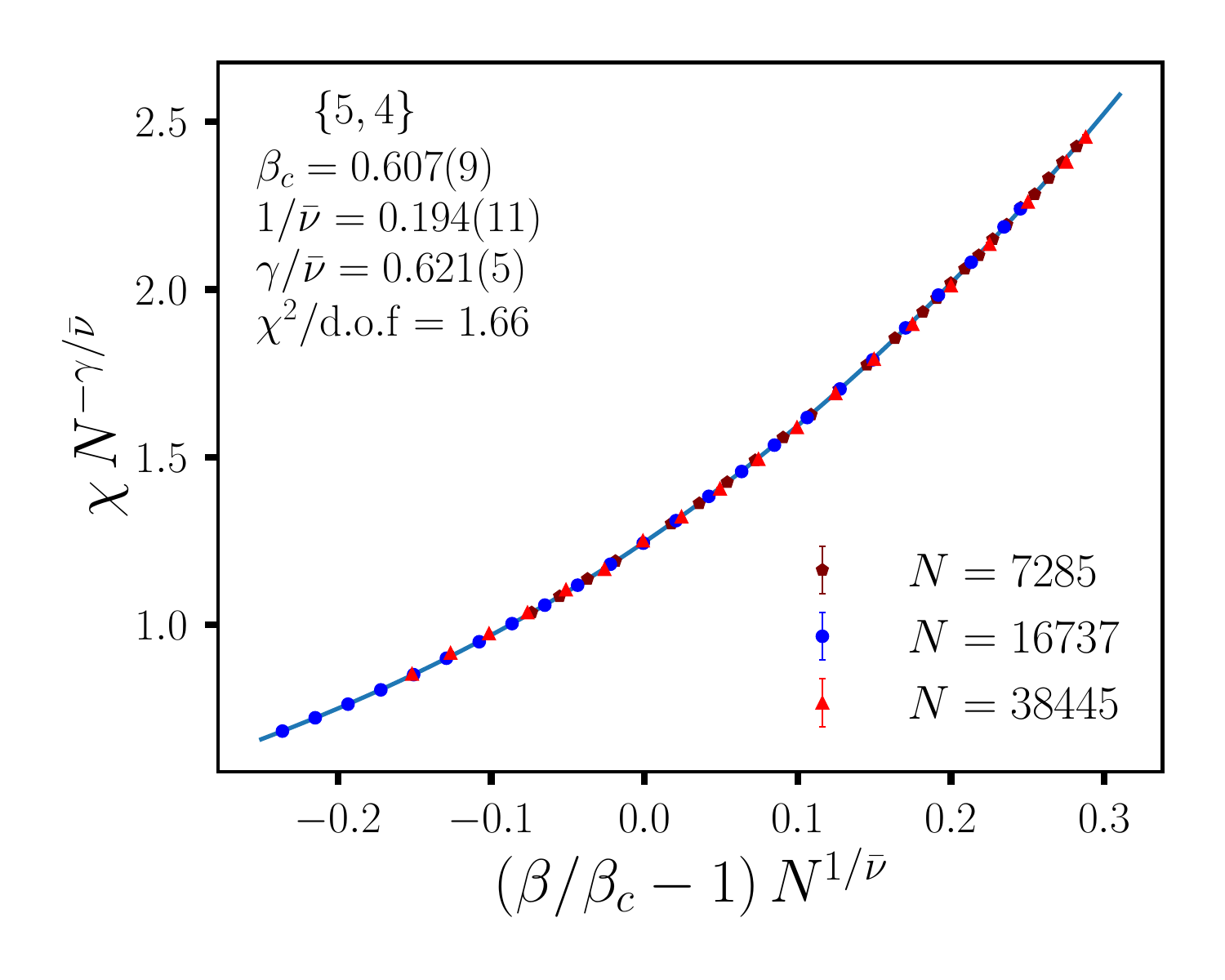}
    \includegraphics[width=0.8\linewidth]{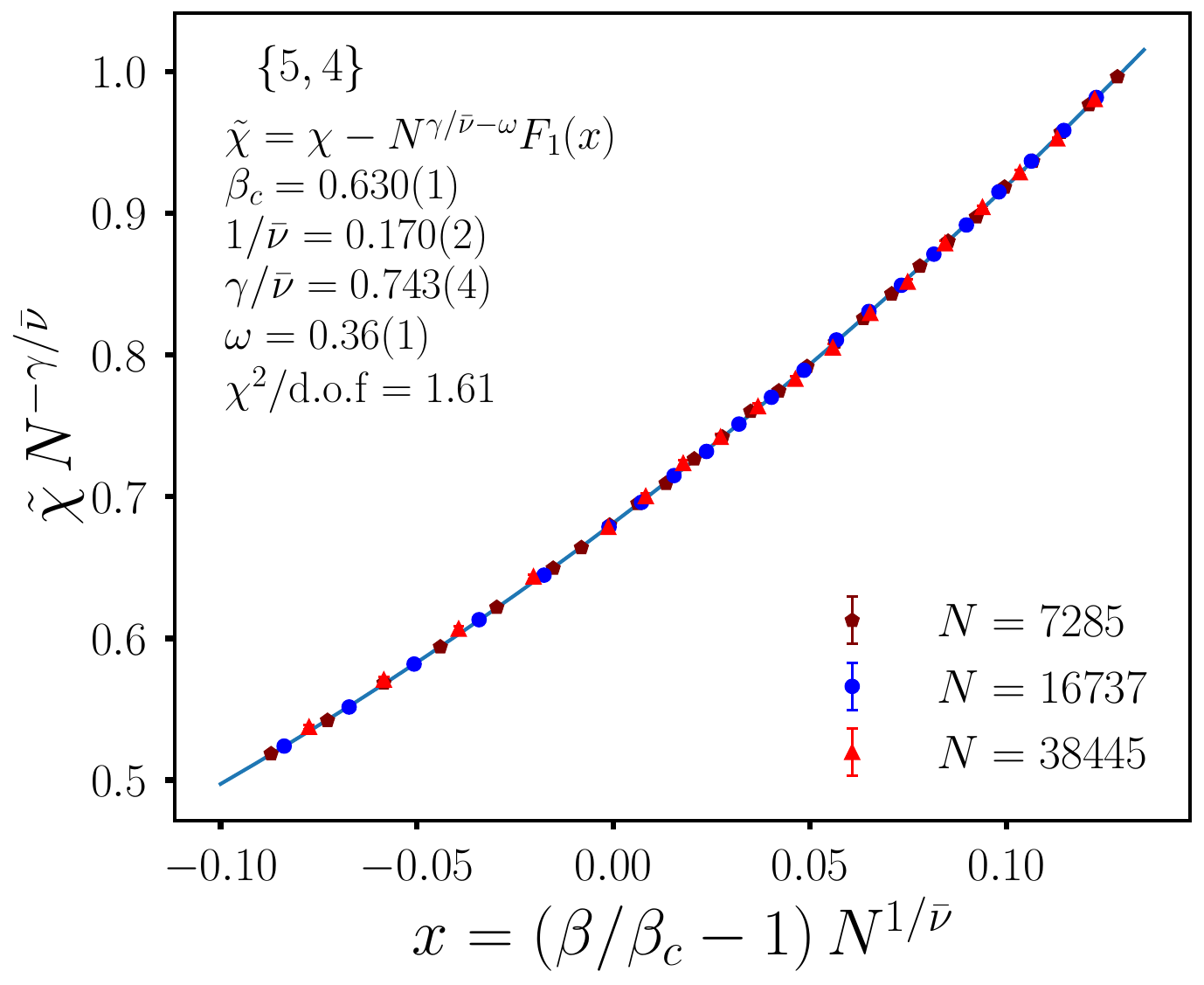}   
    \caption{FSS collapse of $\chi$ for the smaller lattices without (top) and with 
    (bottom) scaling correction term on the $\{5,4\}$ tessellation. The solid line 
    indicates the fourth order polynomial fit and the correction function $F_1(x)$ 
    is the quadratic polynomial in $x = (\beta/\beta_c - 1) N^{1/\bar{\nu}}$.}
    \label{fig:FSS_scalingcorr}
\end{figure}

With OBC, a finite fraction of sites of hyperbolic lattice resides on the boundary, so 
finite-size corrections are expected to be significant. In the main text, we computed
the Binder cumulant $U_2$ using local updates, using layers until $L=12 (N=38445)$, while $\chi$ was extracted using the worm algorithm using layers until $L=13 (N=88301)$. 
Both the crossing point of the Binder cumulant curves (see \cref{fig:crossing_binder}) and 
the FSS (\cref{fig:FSS54}, lower panel) yield $\beta_c = 0.591(2)$.  This (pseudo-)critical 
point is clearly significantly lower than the critical point $\beta_c = 0.632(3)$, obtained 
from the scaling collapse of $\chi$, but using layers until $L=13$, and suggests the 
presence of strong finite size corrections for smaller lattices. Interestingly, if the
FSS with $\chi$ is done to $L=12 (N=38445)$, then we obtain $\beta_c = 0.607(9)$
in closer agreement with the extraction from the Binder cumulant. As would be expected,
using lower layers removes much of the fluctuation, causing the exponents to drift as
well. The results are shown in the top panel of \cref{fig:FSS_scalingcorr}. It is, however,
possible to recover the results of the larger lattices by using an extra correction to
scaling term. As shown in the bottom panel of \cref{fig:FSS_scalingcorr}, with the
correction term, we obtain $\beta_c = 0.630(1)$, $\gamma/\bar{\nu} = 0.743(4)$ and 
$1/\bar{\nu} = 0.170(2)$ are consistent with those from the larger lattices of the 
(top panel) of \cref{fig:FSS54} in the main article and \emph{without correction terms}.
Adding a scaling correction to the FSS of $\chi$ in the largest lattices does not help
to further constrain the fit, and thus we do not show it. 

 While a theoretical study of finite size effects in these lattices is left to a future
study, our cross-checks indicate that the exponents we obtain are not more than 1-2
$\sigma$ away from the ones obtained on even larger lattices.
\subsection{III. Mean field scaling in WBC}
\label{sec:WBC}
\begin{figure*}[t]
    \centering
    \begin{minipage}[t]{0.49\linewidth}
        \centering
        \includegraphics[width=\linewidth]{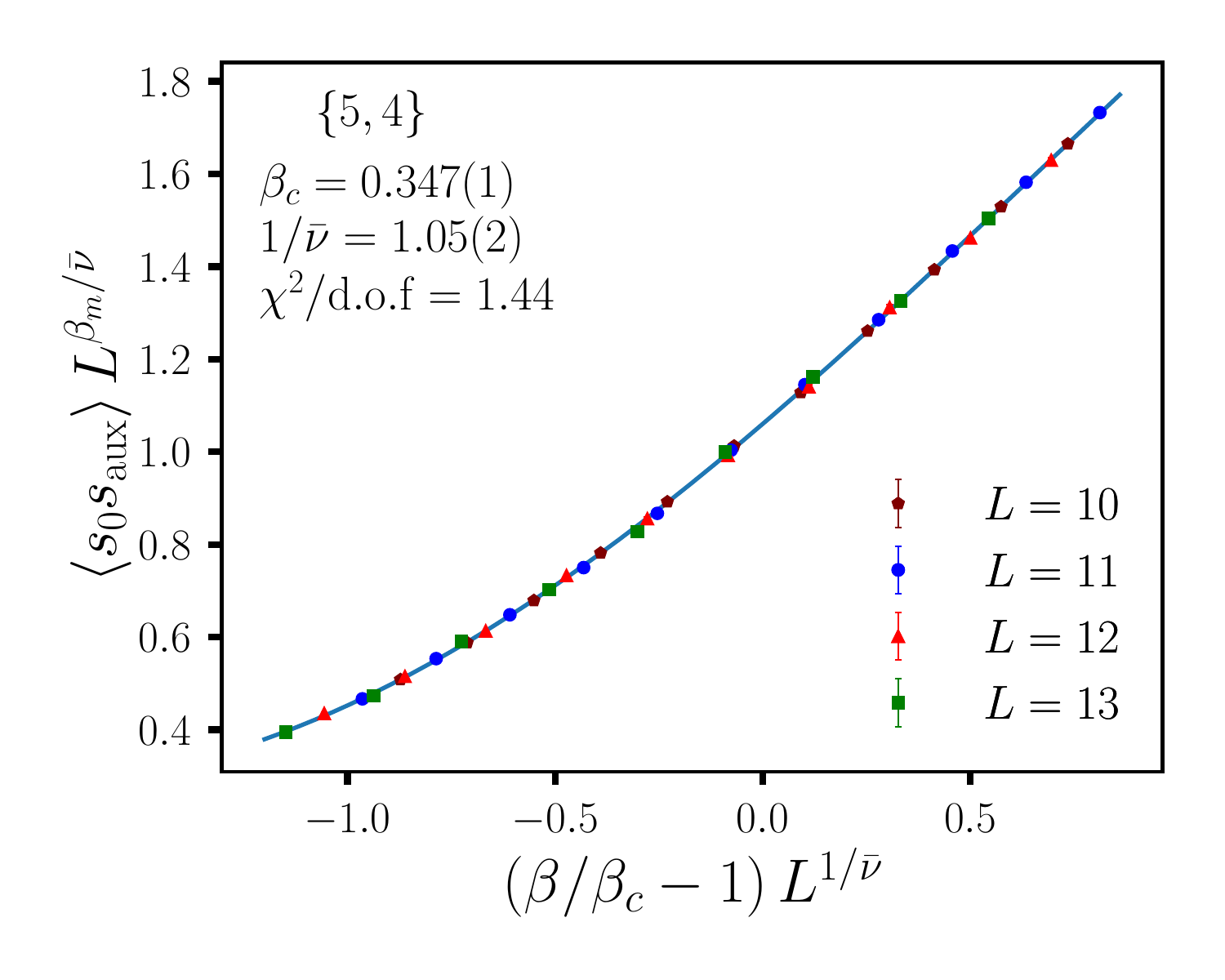}
        \caption{FSS collapse of $\braket{s_0 s_{\rm aux}}$ for the $\{5,4\}$ tiling with WBC, 
        with the exponent fixed to its mean-field value $\beta_m = 1/2$, using a fourth-order 
        scaling polynomial.}
        \label{fig:WBC_fixedMF}
    \end{minipage}\hfill
    \begin{minipage}[t]{0.49\linewidth}
        \centering
        \includegraphics[width=\linewidth]{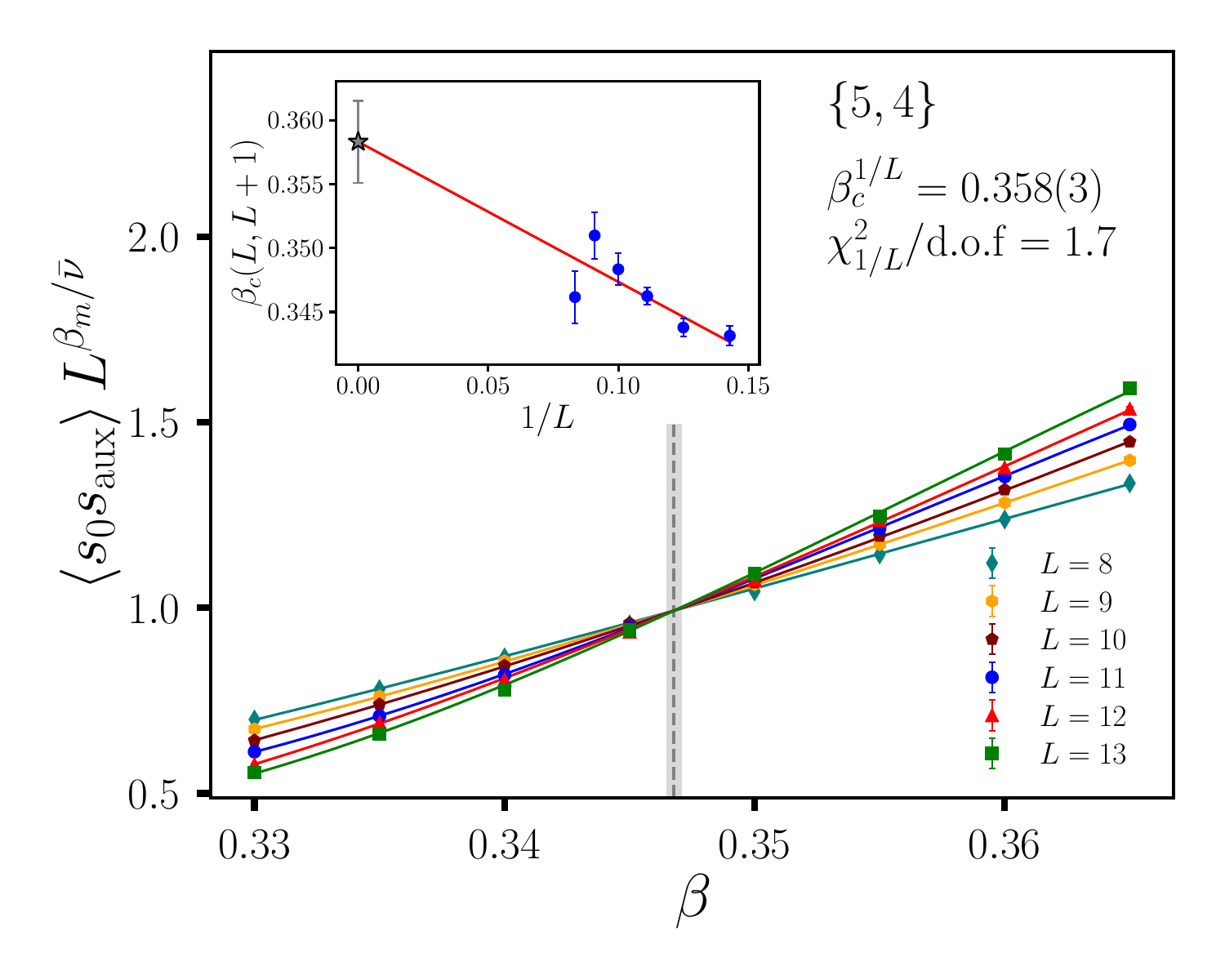}
        \caption{Crossing of the rescaled center-to-boundary correlator
        $\braket{s_0 s_{\rm aux}} L^{\beta_m/\bar\nu}$ with the mean-field value 
        $\beta_m/\bar\nu = 1/2$. Inset: Crossing point of each pair $\beta_c(L,L+1)$ with 
        smallest layer $L$, yields an extrapolated critical point $\beta^{1/L}_c = 0.358(3)$ 
        for the $\{5,4\}$ tessellation.}
        \label{fig:crossing_sosaux}
    \end{minipage}
\end{figure*}
 In this section we offer a different analysis of the WBC data to cross-check the mean field
nature of the phase transition in WBC. In the main text, we performed the FSS by keeping the
exponents as parameters to be fitted. Here, we confirm the mean-field nature of the transition 
with WBC, by fixing the exponent $\beta_m$ to its mean field value, $\beta_m = 0.5$ and repeat 
\cref{fig:tile54_WBC} of the main article for $\braket{s_0 s_{\rm aux}}$ with $\beta_m = 1/2$ 
held fixed. In \cref{fig:WBC_fixedMF}, fixing $\beta_m = 0.5$ yields a good acceptable fit 
with $\beta_c = 0.347(1)$ and $1/\bar{\nu} = 1.05(2)$, both are consistent within the errors 
of \cref{fig:tile54_WBC}. 

To extract the critical point of WBC, we rescale the observable center-to-boundary correlator 
$\braket{s_0s_{\rm aux} }$ by $L^{\beta_m/\bar\nu}$. At criticality, the rescaled correlator
$\braket{s_0 s_{\rm aux}} L^{\beta_m/\bar\nu}$ is scale-invariant, independent of $L$. Using 
the mean-field exponent $\beta_m = 1/2$, the rescaled correlator curve for all lattices indeed 
crosses in narrow region around $\beta_c \approx 0.348$, shown in \cref{fig:crossing_sosaux}. 
The crossing points $\beta_c(L,L+1)$ of a pair of curves $(L, L+1)$ drift slowly to larger 
$\beta$ with increasing $L$ (inset of \cref{fig:crossing_sosaux}). A linear extrapolation 
in $1/L$ gives $\beta_c^{1/L} = 0.358(3)$, in agreement with \cite{f4sj-rwvj}.
\end{document}